\documentclass[aps,a4paper,amsmath,amssymb,twocolumn,
floatfix,groupedaddress]{revtex4}
\usepackage{bm,amsmath,amssymb,amsfonts}
\usepackage[dvips]{graphicx}
\usepackage{color}
\definecolor{dblue}{rgb}{0.0,0.0,0.7}
\definecolor{dred}{rgb}{0.9,0.0,0.0}
\definecolor{dpink}{rgb}{0.85,0.067,0.49}
\usepackage{hyperref}

\hypersetup{
    pdfnewwindow=true,      
    colorlinks=true,       
    linkcolor=dpink,          
    citecolor=blue,        
    filecolor=blue,      
    urlcolor=blue        
}
\newcommand\bea{\begin{eqnarray}}
\newcommand\eea{\end{eqnarray}}
\newcommand\beq{\begin{equation}}
\newcommand\eeq{\end{equation}}

\newcommand{\ie}{{\it i.\,e.}, }

\newcommand{\viz}{{\it viz.}, } 

\begin{document}

\title{Nontrivial topological flat bands in a diamond-octagon lattice geometry}

\author{Biplab Pal} 
\email{biplabpal@pks.mpg.de}

\affiliation{Max Planck Institute for the Physics of Complex Systems, N\"{o}thnitzer Str.\ 38, 
01187 Dresden, Germany }

\date{\today}

\begin{abstract}
We present the appearance of nearly flat band states with nonzero Chern numbers in a two-dimensional 
``diamond-octagon" lattice model comprising two kinds of elementary plaquette geometries, diamond and 
octagon, respectively. We show that the origin of such nontrivial topological nearly flat bands can be 
described by a short-ranged tight-binding Hamiltonian. By considering an additional diagonal hopping 
parameter in the diamond plaquettes along with an externally fine-tuned magnetic flux, it leads to the 
emergence of such nearly flat band states with nonzero Chern numbers for our simple lattice model. 
Such topologically nontrivial nearly flat bands can be very useful to realize the fractional topological 
phenomena in lattice models when the interaction is taken into consideration. In addition, we also 
show that perfect band flattening for certain energy bands, leading to compact localized states can be 
accomplished by fine-tuning the parameters of the Hamiltonian of the system. We compute the density of 
states and the wavefunction amplitude distribution at different lattice sites to corroborate the formation 
of such perfectly flat band states in the energy spectrum. Considering the structural homology between a 
diamond-octagon lattice and a kagome lattice, we strongly believe that one can experimentally realize
a diamond-octagon lattice using ultracold quantum gases in an optical lattice setting. A possible application 
of our lattice model could be to design a photonic lattice using single-mode laser-induced photonic waveguides 
and study the corresponding photonic flat bands. 
\end{abstract}

\maketitle

\section{Introduction}
The physics of flat band (FB) systems has drawn a lot of research attention in recent 
years~\cite{das-sharma-prl2011, tang-prl2011, titus-prl2011, pollmann-prb2015, flach-prb2013, 
flach-epl2014, flach-prl2014, flach-prb2015, flach-prl2016, flach-prb2017, vicencio-pra2017, 
denz-apl2017, ajith-prb2017, ajith-prb2018, biplab-prb2018}. One of the main reasons why 
such dispersionless flat bands are of great interest to the physics community is that, they give rise to 
highly degenerate manifold of single-particle states, which can act as a good platform to study rich, strongly 
correlated phenomena. In a two-dimensional electron gas (2DEG) subject to a strong magnetic field, 
highly degenerate flat Landau levels are formed. It is well-known that completely filled Landau levels 
exhibit integer quantum Hall effect~\cite{klitzing-prl1980} while partially filled Landau levels give rise to 
fractional quantum Hall effect~\cite{laughlin-prb1981}. Generation of nontrivial flat 
bands with nonzero Chern number in 2D tight-binding lattice models may be treated as the lattice 
counterpart of the Landau levels appearing in continuum. Hence occurrence of nontrivial flat bands 
in simple 2D lattice settings can play a pivotal role in investigating profound topological phenomena in 
lattice systems. 

These macroscopically degenerate flat bands with vanishing bandwidth arise in the band structure of 
tight-binding lattice models due to destructive quantum interference of electron hoppings resulting 
in formation of highly localized single-particle states pinned at different atomic sites of the lattice. 
Such highly localized states corresponding to the flat band energies are often attributed to form compact 
localized states (CLS)~\cite{flach-prl2014, flach-prb2017}, modes where the wave function amplitudes 
remain nonzero over a finite number of lattice sites beyond which they sharply decay to zero. These 
flat bands have been found really useful to investigate diverse intriguing phenomena in condensed matter 
physics, \viz ferromagnetism and antiferromagnetism in Hubbard 
models~\cite{tasaki-prl1992, tanaka-prl2003, bitan-prx2014}, superconductivity in 
2D Dirac materials~\cite{kauppila-prb2016}, superfluidity~\cite{peotta-nc2015}, and unconventional Anderson 
localization~\cite{goda-prl2006, shukla-prb2010} are to name a few of them.

Because of the emergence of these important features, flat band systems have been a constant source of 
new ideas to identify novel phenomena involving the interplay between topology and quantum 
physics. On top of that, over the past couple of years some significant experiments featuring flat bands in 
photonic waveguide networks~\cite{vicencio-njp2014, vicencio-prl2015, mukherjee-prl2015, 
mukherjee-ol2015, longhi-ol2014, xia-ol2016, zong-oe2016, weimann-ol2016}, exciton-polariton 
condensates~\cite{masumoto-njp2012, baboux-prl2016}, and ultracold atomic 
condensates~\cite{jo-prl2012, taie-sa2015} have ushered new light into this research domain. 
Spurred by these experimental results, the search for new models with nontrivial 
flat band physics and understanding their usefulness in different lattice geometries have taken a new 
direction in the past few years. In recent times, some important theoretical investigations like the 
role of chiral symmetry on flat bands in a series of tight-binding lattice geometries~\cite{ajith-prb2017}, 
formation of topological flat Wannier-Stark bands in presence of an electric field in a bipartite dice 
network~\cite{ajith-prb2018}, and the emergence of flat bands in fractal-like geometries with various 
interesting band features~\cite{biplab-prb2018} have also enriched the recent literature, revealing 
different subtle issues about flat bands in various lattice geometries.
\begin{figure}[ht]
\includegraphics[clip,width=0.9\columnwidth]{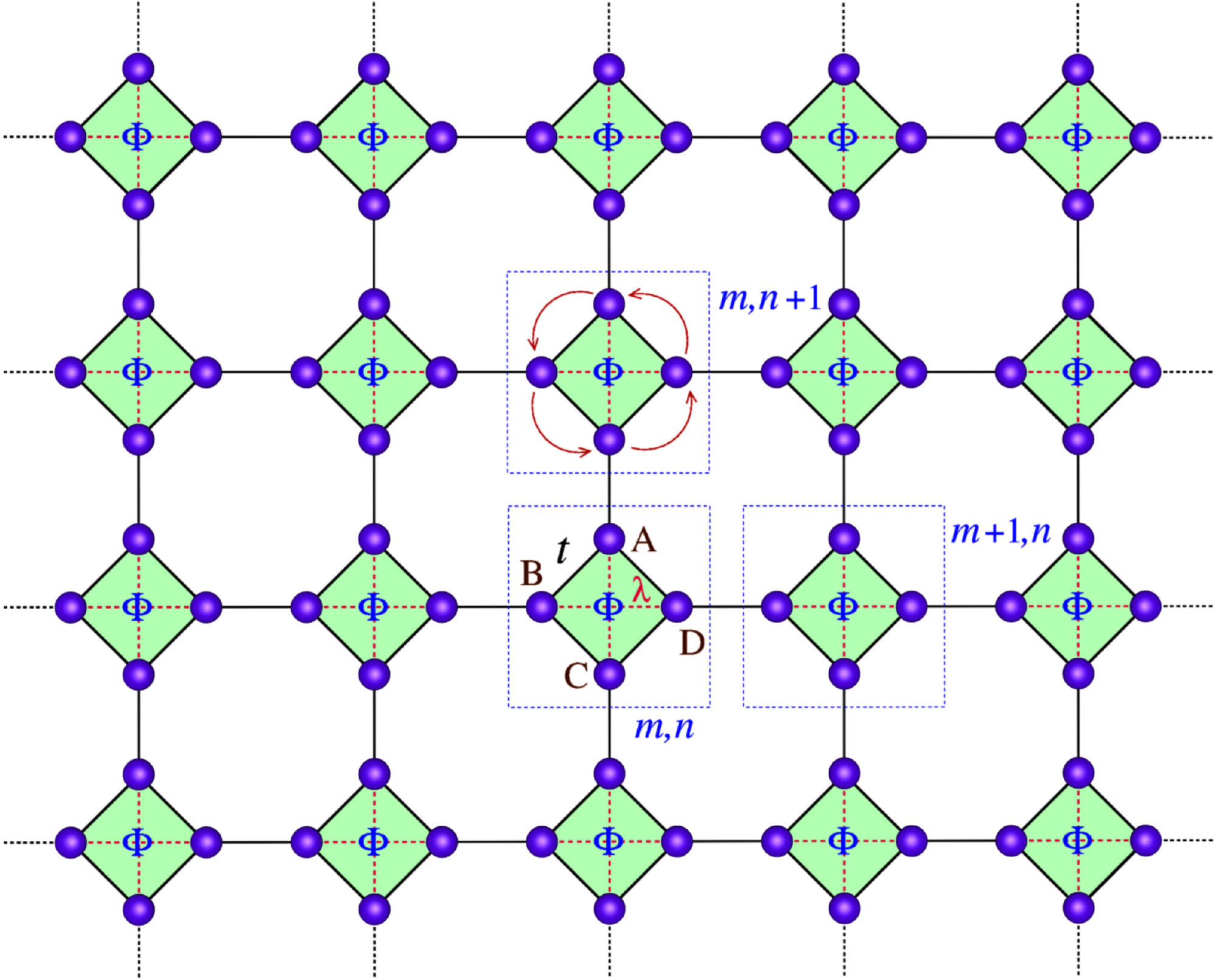}
\caption{Schematic diagram of a 2D diamond-octagon lattice model. The unit 
cells are marked by dotted lines and consist of four atomic sites. The hopping 
parameter along the arms of the diamond and the octagon plaquettes is 
denoted by $t$, and the diagonal hopping integral inside each diamond 
plaquette is represented by $\lambda$. Each diamond plaquette is threaded 
by a uniform external magnetic flux $\Phi$. The arrowheads in the counterclockwise 
direction indicate the direction of the forward hopping in presence of $\Phi$.}
\label{fig:lattice}
\end{figure}

One of the key challenges in generating FB states is to keep the hoppings to be short-ranged. One may 
use the spectral flattening technique, \ie adiabatically transforming the original Hamiltonian to 
a new one with FB states. However, that may often lead to long-range hoppings to be considered in the 
underlying Hamiltonian~\cite{das-sharma-prl2011}, which could be difficult to realize experimentally. 
Other interesting flat band optimization techniques for short-range hopping models have also been 
proposed in recent times~\cite{ronny-prb2016, ronny-prb2017}. It is worth mentioning that, one cannot 
have nontrivial topology, finite-range hopping and exactly flat bands simultaneously -- only two of these 
three criteria can be realized simultaneously~\cite{chen-jpa2014, read-prb2017}. 
In the present paper, we propose  and study a discrete 2D diamond-octagon lattice model 
with short-ranged hopping and an external magnetic flux piercing through the diamond 
plaquettes. This setting leads to the appearance of nontrivial nearly flat band states bearing topological 
properties. Such a lattice model has been incorporated in recent times to study different interesting 
phenomena, such as topological phase transition induced by spin-orbit coupling and non-Abelian gauge 
fields~\cite{fiete-prb2010}, rich magnetic and metal-insulator phases with Hubbard 
interaction~\cite{ueda-prb2013}, and quantum magnetic phase transition with a competitive effect 
between the temperature and the repulsive on-site interaction~\cite{liu-sr2014}. 

In this work we analyze the band spectrum of diamond-octagon lattice in a single particle picture. This 
lattice geometry realizes a four-band model in the momentum space. Each diamond-shaped loop is pierced by a 
uniform external magnetic flux which breaks the time-reversal symmetry appending an Aharonov-Bohm 
phase~\cite{aharonov-pr1959} to the hopping parameter along the arms of each diamond plaquette. This leads 
to gapping out of the band spectrum with bands having nonzero Chern numbers. We furthermore 
show that one can fine-tune the hopping parameters along with a suitable value of the magnetic flux to 
achieve the optimal band flatness for the bands having the nontrivial topological index in the form of nonzero 
Chern number. It has been argued previously by other groups that perfect flatness for a band in real materials is 
not a stringent requirement provided that the bandwidth remains much smaller than the band 
gap~\cite{das-sharma-prl2011, tang-prl2011, titus-prl2011}. 
These nearly flat bands having strong resemblance with the Landau levels appearing in a continuum 2DEG model, 
set up a good foundation to explore new strongly correlated topological states of matter. We note that such 
tight-binding lattice models with a variety of lattice geometries such as 
Lieb~\cite{manninen-pra2010, goldman-pra2011}, 
kagome~\cite{altman-prb2010}, honeycomb~\cite{wu-prl2007}, square~\cite{monika-np2015}, etc. have 
been proposed to be realized using ultracold fermionic or bosonic atoms in optical lattices. 

In what follows, we give an illustration of  our model and present the important findings. In 
Sec.~\ref{model}, we introduce our lattice model and discuss the short-ranged tight-binding 
Hamiltonian describing the spinless particles moving on the lattice. In Sec.~\ref{tfb}, we 
discuss the condition for generating the nearly flat bands in the band spectrum, 
and present the results for the Berry curvature and the Chern numbers corresponding to the nontrivial 
topological flat bands. This is followed by Sec.~\ref{cfb}, where we describe how to create perfect flat 
bands in the band structure by tuning the combination of hopping parameters and the magnetic flux. 
We also compute the average density of states and the wavefunction amplitude distribution on different 
lattice sites corresponding to such perfect flat band states. In Sec.~\ref{expt}, we depict the scope of 
a possible experimental set up using single-mode photonic waveguide structure to realize our lattice model in an 
actual experiment. Finally, in Sec.~\ref{summary}, we draw our conclusion with a summary of our results 
and their utility with the scope of future study in this direction. 

\section{The model and the mathematical framework}
\label{model}
We consider a diamond-octagon lattice model on a two-dimensional plane comprising two elementary plaquette 
geometries, \viz diamond and octagon, respectively as shown in Fig.~\ref{fig:lattice}. The building block of 
the lattice structure is a diamond-shaped loop consisting of four atomic sites. This basic unit cell is repeated 
periodically over a two-dimensional plane to form the whole lattice structure. Each diamond plaquette is pierced by 
a uniform magnetic flux perpendicular to the plane of lattice which introduces an Aharonov-Bohm phase to the 
hopping parameter when an electron hops along the boundary of a diamond loop. The tight-binding Hamiltonian 
of this model in Wannier basis can be written as,
\begin{equation}
\bm{H} = \sum_{m,n}\Big[ \sum_{i}\epsilon_{i} c_{m,n,i}^{\dagger}c_{m,n,i}\Big] + 
\Big[ \sum_{i,j}\mathcal{T}_{ij} c_{m,n,i}^{\dagger}c_{m,n,j} + \textrm{H.c.} \Big],
\label{eq:hamil-wannier}
\end{equation}
where the first summation runs over the unit cell index $(m,n)$ as shown in Fig.~\ref{fig:lattice}. 
$c_{m,n,i}^{\dagger}$ $(c_{m,n,i})$ is the creation (annihilation) operator for an electron at site $i$ in the 
$(m,n)$-th unit cell and $\epsilon_{i}$ is the on-site potential for the $i$-th atomic site. $\mathcal{T}_{ij}$ 
is the hopping parameter between the $i$-th and the $j$-th sites, and it can take two possible values depending 
on the position of the sites $i$ and $j$. $\mathcal{T}_{ij} = t$ for an electron hopping along the boundary 
of a diamond plaquette or along the line in between two consecutive diamond loops, 
and $\mathcal{T}_{ij} = \lambda$ for an electron hopping along the diagonals inside a diamond plaquette. 
Each diamond plaquette is pierced by an external magnetic flux $\Phi$ which incorporates an Aharonov-Bohm 
phase factor to hopping parameter $t \rightarrow t \exp{(\pm i\Theta)}$, when the electron hops around the 
closed loop in a diamond plaquette. Here, $\Theta = 2 \pi \Phi / 4 \Phi_{0}$, $\Phi_{0}=hc/e$ being the 
fundamental flux quantum, and the sign $\pm$ in the exponent indicates the direction of the forward and the 
backward hoppings. 

By adopting a momentum $(\bm{k})$ space description using a discrete Fourier transform, the Hamiltonian in 
Eq.~\eqref{eq:hamil-wannier} can be recast as,
\begin{equation}
\bm{H} = \sum_{\bm{k}} \bm{\Psi}^{\dagger}_{\bm{k}} \bm{\mathcal{H}}(\bm{k})\bm{\Psi}_{\bm{k}},
\label{eq:hamil-mom} 
\end{equation}
where $
\bm{\Psi}^{\dagger}_{\bm{k}} \equiv
\left(\begin{matrix}
c^{\dagger}_{k_x,k_y,A}  &  c^{\dagger}_{k_x,k_y,B}  
&  c^{\dagger}_{k_x,k_y,C} & c^{\dagger}_{k_x,k_y,D} 
\end{matrix}\right)$,
and $\bm{\mathcal{H}}(\bm{k})$ is given by,
\begin{align}
\bm{\mathcal{H}}({\bm k}) =
\left(\def\arraystretch{1.5} \begin{matrix}
0  &  te^{i\Theta}  &  te^{ik_y}+\lambda  &  te^{-i\Theta} \\
te^{-i\Theta}  &  0  &  te^{i\Theta}  &  te^{-ik_x}+\lambda \\
te^{-ik_y}+\lambda  &  te^{-i\Theta}  &  0  &  te^{i\Theta} \\
te^{i\Theta}  &  te^{ik_x}+\lambda  &  te^{-i\Theta}  &  0 \\ 
\end{matrix}\right).
\label{eq:ham}
\end{align}
We have taken $\epsilon_{i}=0$, $i\in \{A,B,C,D\}$. One can extract all the interesting features about 
the band structure of the system from Eq.~\eqref{eq:ham}. The results are presented in the next section. 
\section{Generation of nearly flat topological bands}
\label{tfb}
The standard prescription for investigating any special feature of a lattice model is to frame the tight-binding 
Hamiltonian in $\bm{k}$-space, and then minutely study its band structure by playing around the parameters of 
Hamiltonian, namely, short-ranged hopping strengths, or some external perturbations like magnetic field, 
electric field, disorder, etc. These effects often lead to some interesting topological properties in simple 
tight-binding lattice models. In the present study we embark on such a lattice geometry. Our aim is to discover 
whether this lattice structure can show up some nontrivial topological properties in its band structure under 
certain special condition of the parameters space of the corresponding Hamiltonian. 
\begin{figure}[ht]
\includegraphics[clip,width=0.8\columnwidth]{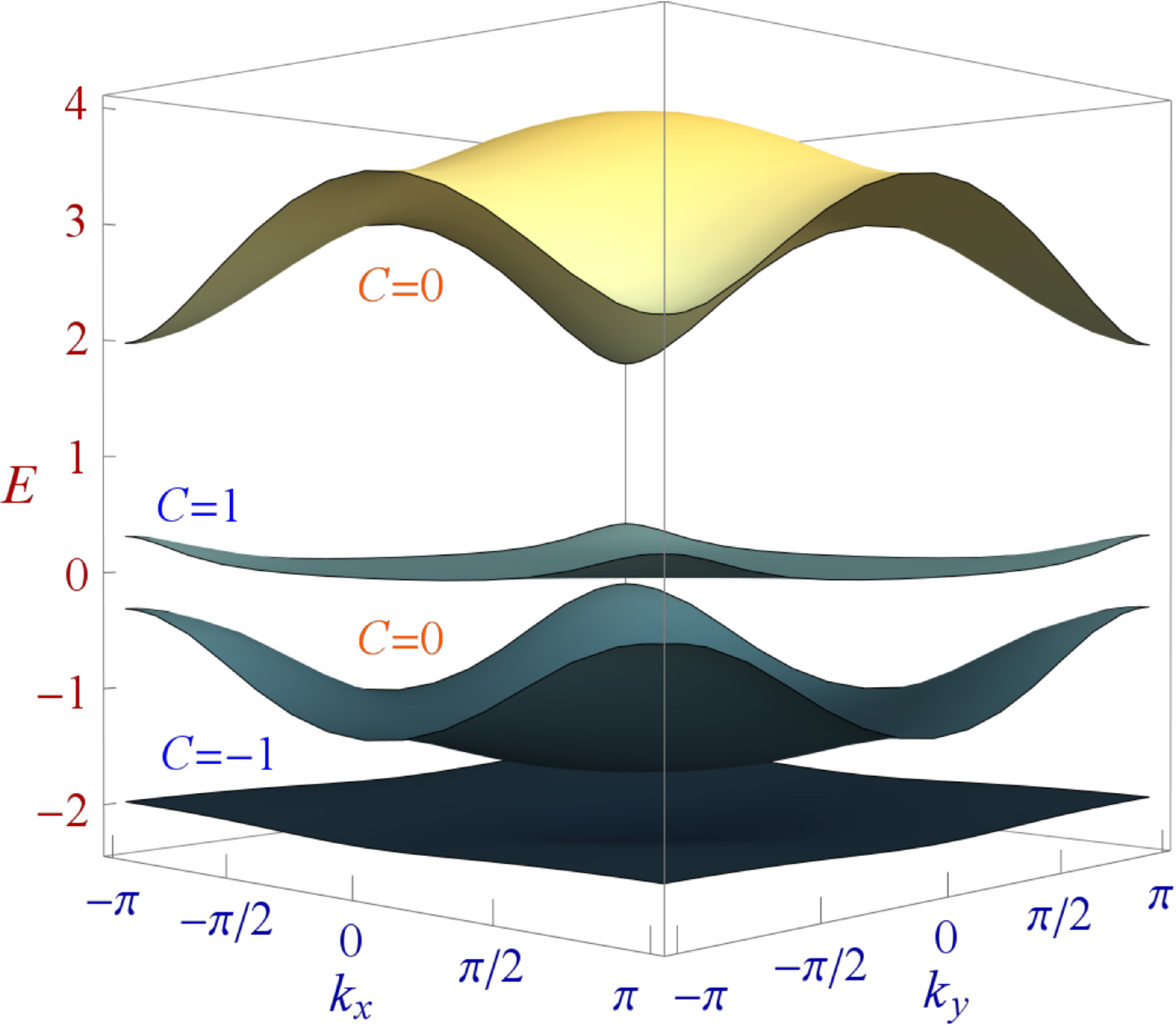}
\caption{Plot of the band structure for the 2D diamond-octagon lattice model 
prescribed in Fig.~\ref{fig:lattice}. The lowest and the third band show nontrivial 
topological character with nonzero integer values of the Chern numbers, \viz $C=-1$ and 
$C=1$, while the remaining two bands are topologically trivial with zero Chern numbers. 
We tune a minimal nonzero value of the external magnetic flux $\Phi=\Phi_{0}/10$, and the 
short-ranged hopping parameters are set to be $t=1$ and $\lambda=1$, respectively. 
These are the optimized values of the parameters to achieve the optimized flatness of the 
two topologically nontrivial bands.}
\label{fig:tfb}
\end{figure}

The Hamiltonian in Eq.~\eqref{eq:ham} describes our model. It is apparent from Eq.~\eqref{eq:ham} that the Hamiltonian 
breaks the time-reversal symmetry for a nonzero value of the Aharanov-Bohm phase $\Theta$, which implies to have 
$\Phi \neq 0$. Breaking of such time-reversal symmetry in the system by introduction of a complex phase factor 
in the hopping parameter through a staggered magnetic flux~\cite{nagaosa-prb2000, green-prb2010} or through 
some artificial gauge field~\cite{das-sharma-prl2011, haldane-prl1988} have insightful consequences 
on the band structure as well as to the topological properties of the system as evinced in other previous 
important studies. At $\Phi \neq 0$, band gap opens up in between different bands of our system. In presence of 
such gap opening in the system, we have calculated the Chern numbers corresponding to different bands of 
the system, and discovered that two of the bands possess nonzero Chern numbers indicating nontrivial topological 
character of those bands. The scenario in which we are interested in, is to have the optimized flatness of the bands 
carrying nonzero Chern numbers. To acquire such a condition, we optimize the values of parameters of the Hamiltonian 
such as the hopping strengths and the external magnetic flux. It turns out that for the optimized nearly 
flat bands with nonzero Chern numbers, the values of the parameters are found to be $\Phi=\Phi_{0}/10$, 
$t=1$, and $\lambda=1$. The flatness ratio~\cite{ronny-prb2016, ronny-prb2017} for the Chern bands 
with these model parameters is approximately 5. 
We first numerically evaluate the values of the hopping integrals $t$ and $\lambda$ for which we have 
perfectly flat bands in the spectrum in absence of any magnetic flux. Then we tune the magnetic flux 
to a nonzero value to have the gap opening in the band spectrum. Under this condition, we fix the magnetic 
flux to a value where we get the maximum gap to bandwidth ratio for the Chern bands. 
We note that for a minimal deviation from these particular values of the 
model parameters, the band features will not alter a lot. Thus, there is a realistic possibility to realize our 
results in an actual experimental situation where one needs a little bit of relaxation on the exact 
conditions of the parameters. 

\begin{figure}[ht]
\includegraphics[clip,width=0.49\columnwidth]{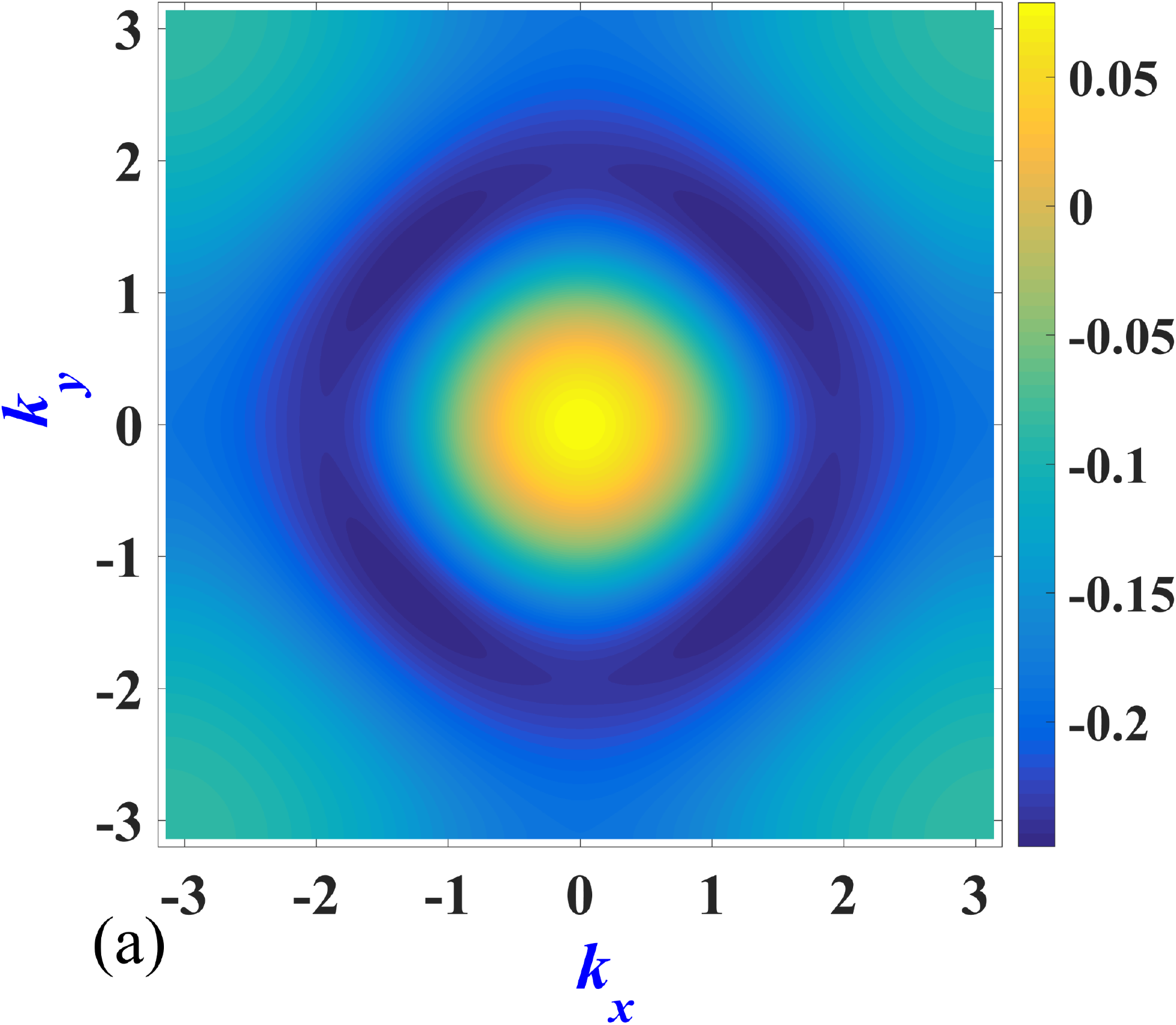}
\includegraphics[clip,width=0.49\columnwidth]{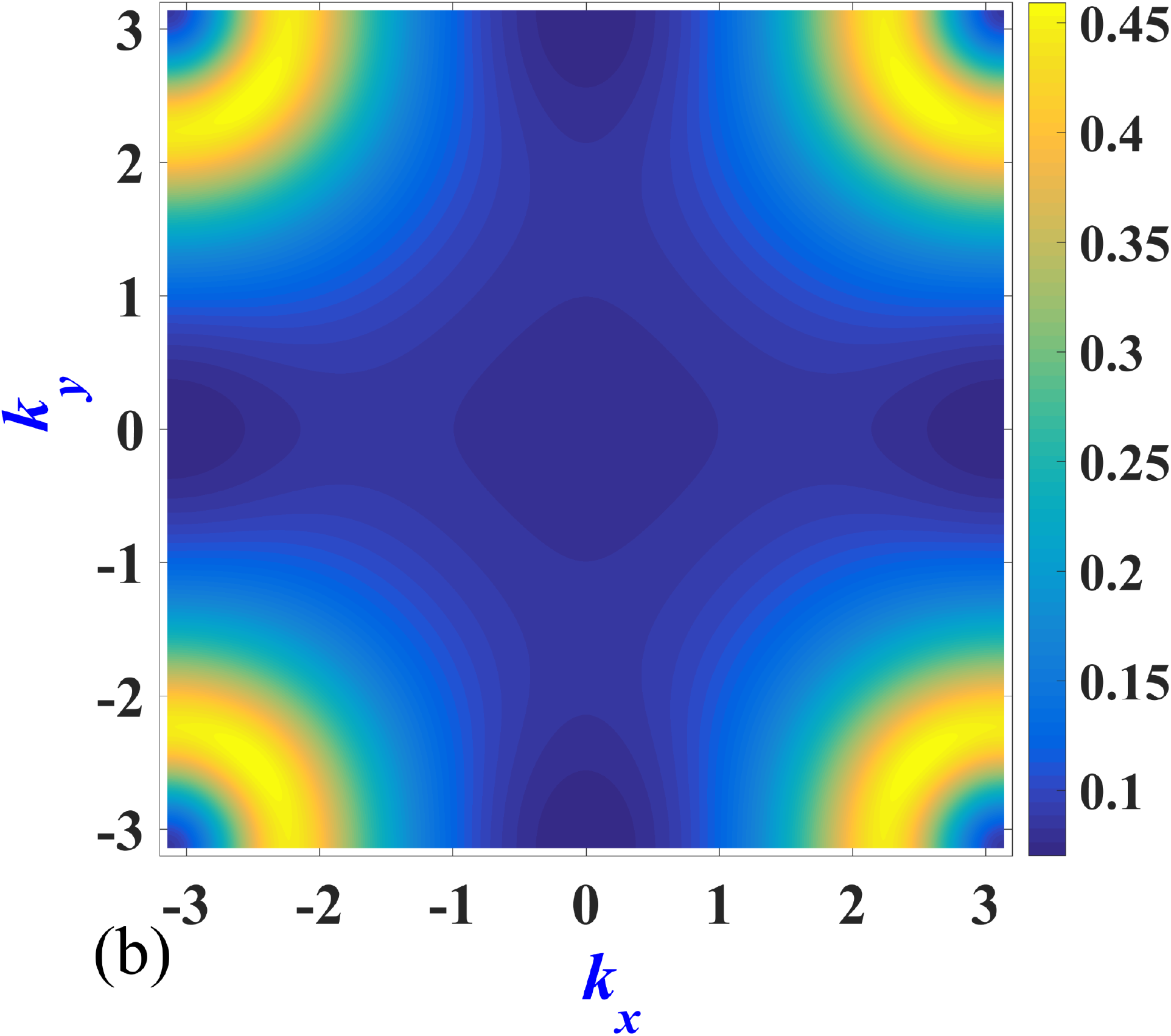}
\caption{Plot of the Berry curvature distribution in momentum space corresponding to 
the topologically nontrivial bands carrying integer Chern numbers. Panel (a) is for the 
lowest band ($n=1$) with $C=-1$ and panel (b) is for the third band ($n=3$) with $C=1$.}
\label{fig:BC}
\end{figure}

To study the topological properties of the bands we calculate the Berry curvature of all the bands 
using the standard formula~\cite{haldane-prl2004, chen-jpcm2012} given by,
\begin{widetext}
\begin{equation}
\Omega_{n} (\bm{k}) = \sum_{m \neq n} \dfrac{-2\textrm{Im}\left[ \langle \mathcal{U}_{n}(\bm{k}) | 
\partial \bm{\mathcal{H}}(\bm{k})/ \partial k_{x} | \mathcal{U}_{m}(\bm{k}) \rangle 
\langle \mathcal{U}_{m}(\bm{k}) | 
\partial \bm{\mathcal{H}}(\bm{k})/ \partial k_{y} | \mathcal{U}_{n}(\bm{k}) \rangle \right]} 
{(E_{n}-E_{m})^2},
\label{eq:BC}
\end{equation}
\end{widetext}
where $\mathcal{U}_{n}(\bm{k})$ is the $n$-th eigenstate of  $\bm{\mathcal{H}}(\bm{k})$ with an 
energy eigenvalue $E_{n}(\bm{k})$. Using Eq.~\eqref{eq:BC}, one can easily evaluate the value of the Chern number 
for each of the bands of the system using the following expression,
\begin{equation}
C = \dfrac{1}{2\pi}\int_{BZ} \Omega_{n} (\bm{k}) d\bm{k},
\label{eq:Chern}
\end{equation} 
where $BZ$ stands for the first Brillouin zone of the corresponding lattice structure. We have taken the lattice 
constant $a$ to be unity throughout our calculations. Using the above prescription, we have discovered that in 
our four-band lattice model, two of the bands, \viz the third and the lowest one possess nonzero Chern numbers 
$C=\pm 1$ exhibiting the topological character, and the remaining two bands are topologically trivial with $C=0$ 
as indicated in Fig.~\ref{fig:tfb}. We show the variations for the Berry curvature in the momentum space 
corresponding to the topologically nontrivial bands in Fig.~\ref{fig:BC}.
One can easily observe the distinct features appearing in the Berry curvature distributions for the 
topologically nontrivial bands as apparent from Figs.~\ref{fig:BC}(a) and~\ref{fig:BC}(b). 

One of the most exciting features about our result is that the bands having nonzero Chern numbers are 
nearly flat. Such nearly flat bands with nonzero Chern numbers can be thought of as the lattice 
analogue of the Landau levels appearing in a continuum system. Thus, at fractional filling, our 
lattice model can act as a potential setting to investigate and understand the fractional quantum 
Hall physics in a lattice model with the interactions between the particles being treated as 
just subleading corrections. We note that the inclusion of the diagonal hopping $\lambda$ in 
between the sites inside the diamond plaquette in our model plays an important role in 
generating the flat bands for our system. Such diagonal hopping parameters in between the 
lattice sites have been taken into consideration previously for some other interesting 
tight-binding lattice models~\cite{das-sharma-prl2011, biplab-prb2012}. One can easily 
verify that, in absence of $\lambda$, the band structure of our lattice model will give rise to 
two interpenetrating conical shapes as shown in Ref.~\cite{liu-sr2014}. We also note that some 
interesting topological quantum phase transitions have been reported earlier~\cite{liu-jpcm2013} 
on the similar lattice structures considering a different Hamiltonian with up to third 
nearest neighbor hopping parameter. They have discussed the possibility of having 
topological approximate flat bands as well as higher Chern numbers in the 
system for certain other parameter regimes. However, in our model we only consider short-ranged 
hopping parameters to show the topological properties of the band structure. In addition to 
that, we have discovered that perfect band flattening can also be achieved for our model for 
certain combinations of the parameter values. This is discussed in detail in the next section.
\section{Formation of perfect flat bands}
\label{cfb}
The focus of this section is to explore the conditions for obtaining the complete FB states for our lattice 
system. In tight-binding lattice models often the interplay between the lattice topology and the destructive quantum 
interference among the particle hoppings lead to formation of perfect FB states. The particles in these 
states do not hop to the neighboring lattice sites and form highly localized states. The effective mass of 
the particles in such situation can be thus viewed as infinite. This phenomenon can appear both in absence and 
in presence of broken time-reversal symmetry in the lattice systems~\cite{nagaosa-prb2000, green-prb2010}. 
\begin{figure}[ht]
\includegraphics[clip,width=0.49\columnwidth]{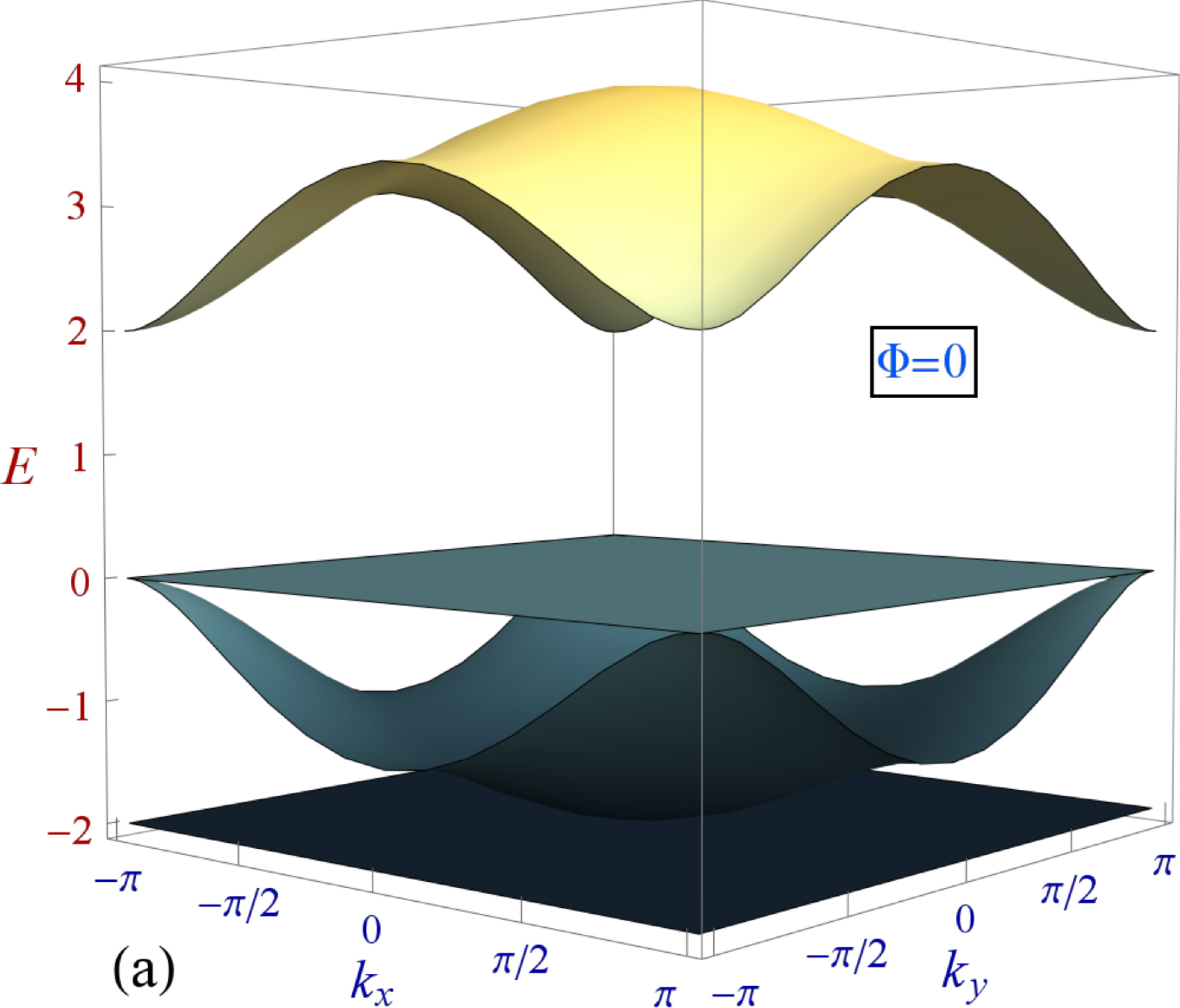}
\includegraphics[clip,width=0.49\columnwidth]{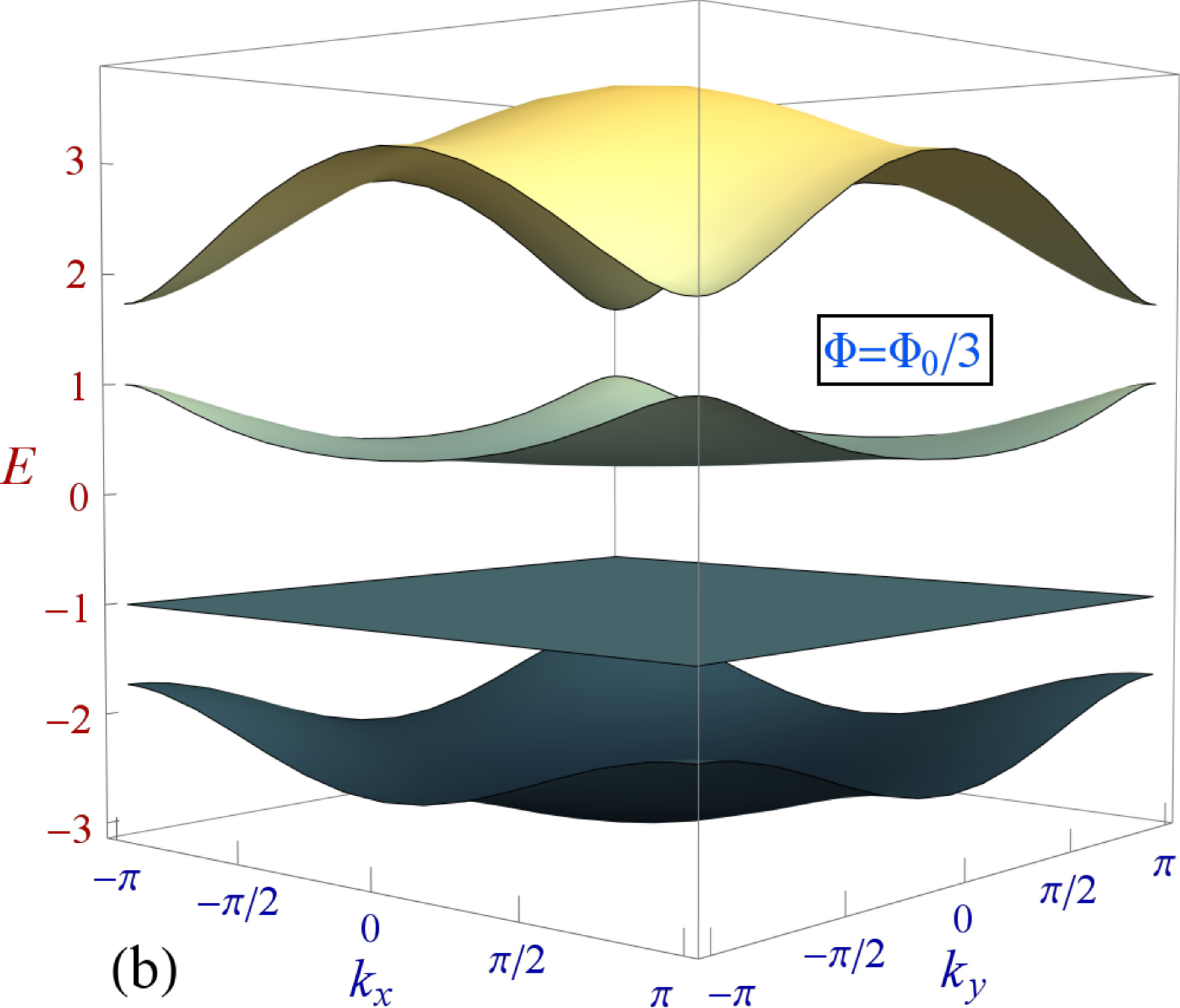}
\caption{Formation of complete flat bands both in absence and in presence of the 
external magnetic flux $\Phi$. The left panel is for $\Phi=0$ and the right panel 
corresponds to $\Phi=\Phi_{0}/3$. We set $t=1$ and $\lambda=1$.}
\label{fig:cfb}
\end{figure}

\begin{figure}[ht]
\includegraphics[clip,width=0.9\columnwidth]{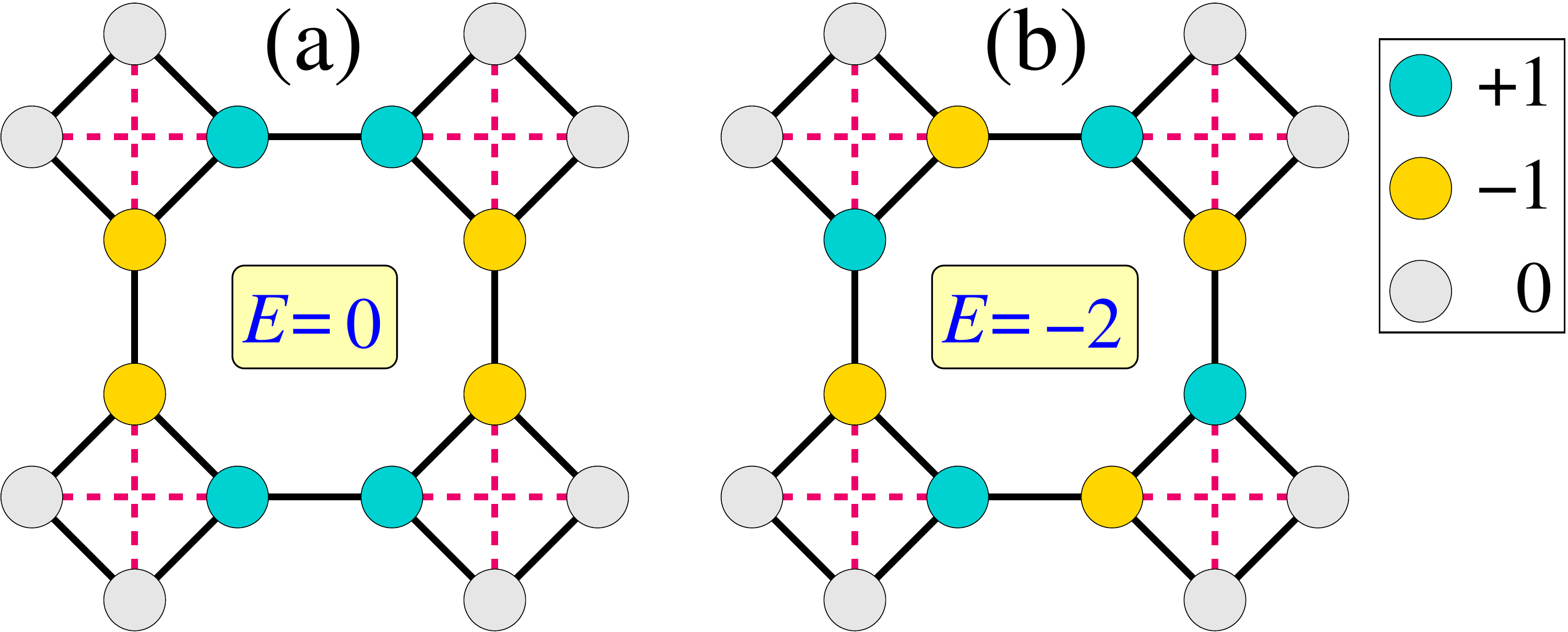}
\caption{Distribution of wavefunction amplitudes at different lattice sites for the 
compact localized states corresponding to the FB states with energies (a) $E=0$ 
and (b) $E=-2$. The on-site energy for all the sites is set to zero, and the values 
of the other parameters are $\Phi=0$, $t=1$, and $\lambda=1$. The values of 
the wavefunction amplitudes at different lattice sites are $+1$, $-1$, and $0$, 
respectively.}
\label{fig:amp-distri}
\end{figure}

First we analyze the case with $\Phi=0$, \ie in absence of time-reversal symmetry breaking. For $\Phi=0$, 
our lattice model yields two perfectly flat dispersionless bands in the band spectrum at energies 
$E_{\mathrm{FB}}=0$ and $-2t$, respectively, as shown in Fig.~\ref{fig:cfb}(a). The values of the 
hopping parameters for these flat bands are $t=1$ and $\lambda=1$, respectively. The 
two flat bands are accompanied by two dispersive bands in the spectrum, one of which is completely isolated from 
the rest of the bands and the other one is sandwiched in between the two perfect flat bands. This is in marked 
contrast with the frustrated hopping models, in which the dispersionless energy band occurs only at the maximum 
or minimum of the spectrum in absence of any magnetic field~\cite{nagaosa-prb2000, balents-prb2008}. Such 
flat bands appearing in absence of any magnetic flux allow for the formation of compact localized 
states (CLS)~\cite{flach-prb2017, ajith-prb2017}. In CLS, the compact eigenstates are perfectly localized over 
a few lattice sites, with exactly vanishing wavefunction amplitudes on all other sites~\cite{ajith-prb2017, 
biplab-prb2018}. Using a standard technique~\cite{biplab-prb2018}, we have worked out the distribution 
of wavefunction amplitudes at different lattice sites corresponding to the CLS for our model. The results are 
presented in Fig.~\ref{fig:amp-distri}. It is clear from Figs.~\ref{fig:amp-distri}(a) and~\ref{fig:amp-distri}(b) 
that the wavefunctions corresponding to the FB states are localized over a couple of lattice sites with nonzero 
amplitudes (marked by dark colored circles), and beyond that the wavefunction amplitudes decay to zero (marked 
by light gray circles). We note that, the perfect flat bands for $\Phi=0$ can be attributed to the fact that the 
diamond-octagon lattice is a line graph of the Lieb lattice~\cite{mielke-jpa1991}.

Next we consider the scenario with $\Phi \neq 0$, which breaks the time-reversal symmetry for our model. 
In presence of the time-reversal symmetry breaking, the bands are gapped out. It turns out that for a nonzero value of 
the magnetic flux between $0$ and $\Phi_{0}$, the FB states get destroyed, leading to dispersive bands in the band spectrum 
with opening of gaps in between them. However, we have discovered that for certain special values of the magnetic flux 
$\Phi$, the FB states re-emerge in the spectrum. For example, an isolated completely FB state emerges at the energy 
$E_{\mathrm{FB}}=-t $ in the spectrum for a value of the magnetic flux $\Phi = \Phi_{0}/3$ as depicted 
in Fig.~\ref{fig:cfb}(b). The values of the hopping integrals are $t=1$ and $\lambda=1$, respectively. 
The resulting FB ($n=2$), however, turns out to be topologically trivial with zero Chern 
number while the dispersive bands, namely, the lowest ($n=1$) and the third ($n=3$) one show up topological 
character with nonzero integer values of the Chern numbers, \viz $C=-1$ and $+1$, respectively. We note that 
this result is consistent with the previous interesting studies on other similar tight-binding lattice models 
such as kagome or hexagonal lattice~\cite{green-prb2010} and Lieb lattice~\cite{franz-prb2010}. One can also 
have similar situation for $\Phi = 2\Phi_{0}/3$. We note that, for $\Phi=0$ we have the gapless perfectly 
flat bands in the spectrum, and as we tune $\Phi$ to a nonzero value, there is a gap opening and appearance of nearly flat 
bands in the spectrum with nonzero Chern numbers. So we have a clear transition from perfectly flat band states to nearly flat 
Chern bands as we change the magnetic flux $\Phi$ from a zero to a nonzero value.
\begin{figure}[ht]
\includegraphics[clip,width=0.48\columnwidth]{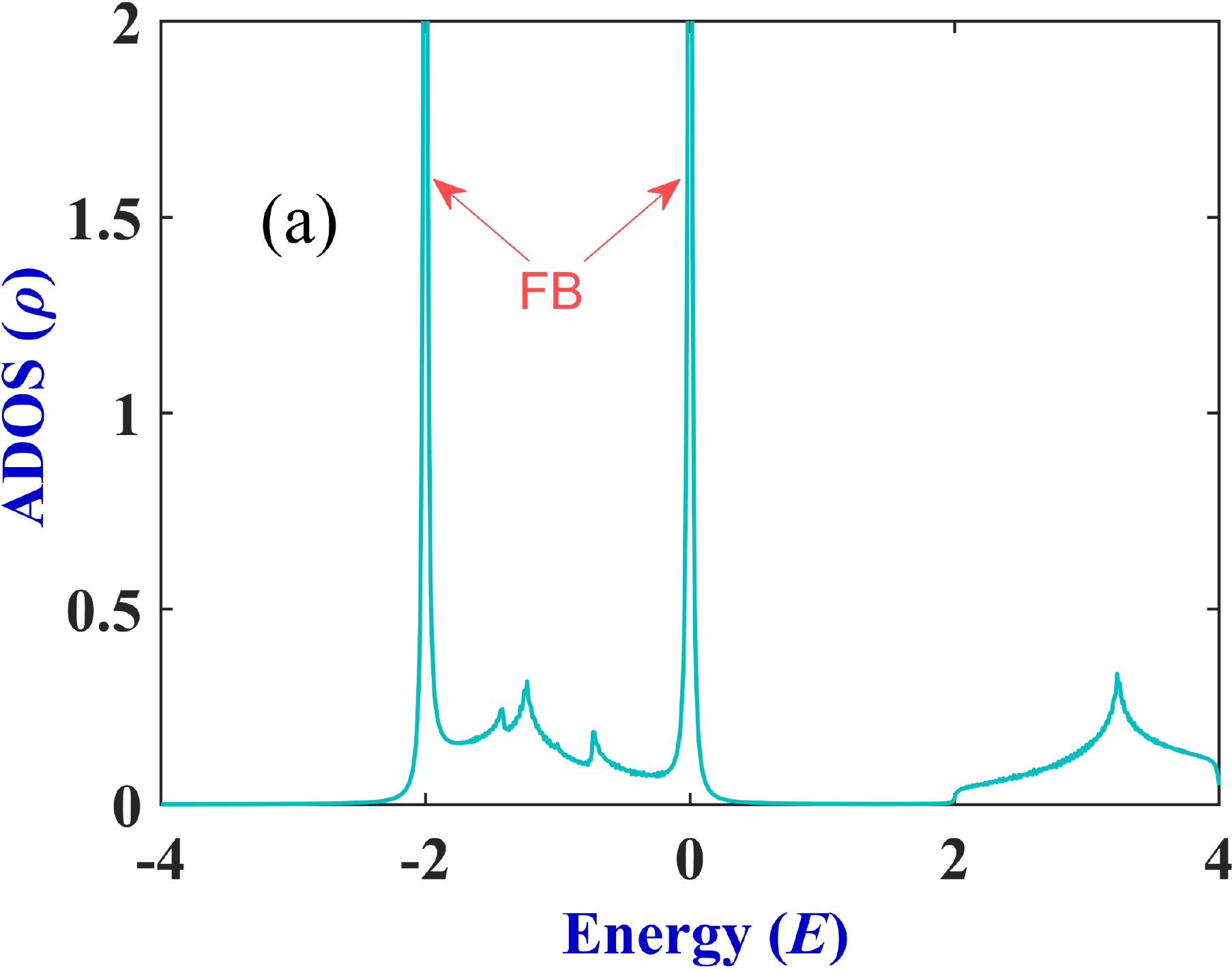}
\includegraphics[clip,width=0.48\columnwidth]{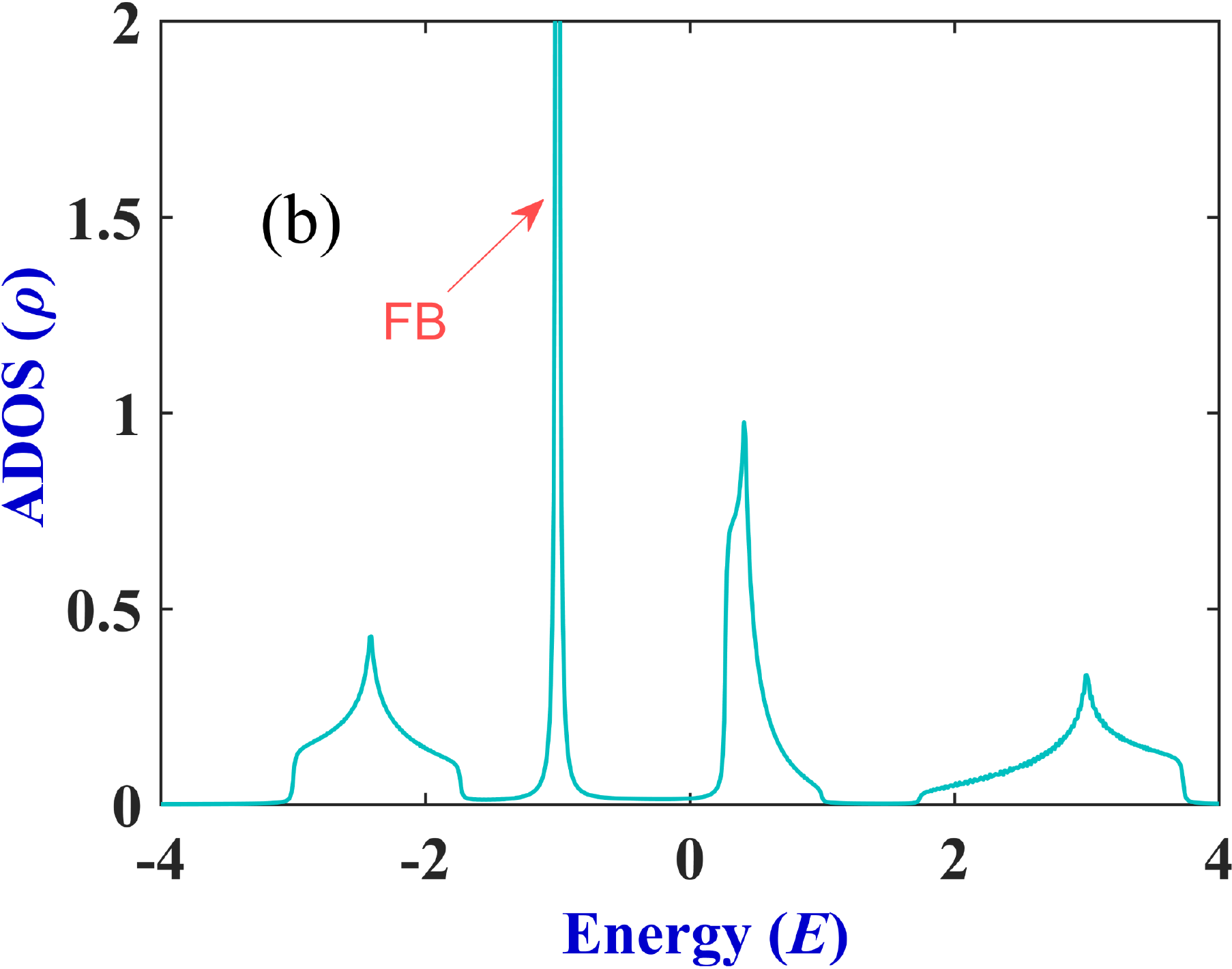}
\caption{Plot of the average density of states (ADOS) for a 2D diamond-octagon 
lattice structure with system size  $L=50 \times 50$. For panel (a) we have $\Phi=0$ and 
for panel (b) we set $\Phi=\Phi_{0}/3$. The other parameters are same as in Fig.~\ref{fig:cfb}. 
The FB states in the ADOS spectrum are indicated by red arrowheads.}
\label{fig:ADOS}
\end{figure}

To substantiate the fact that the particles dwelling in a complete FB state are highly localized, we compute the 
average density of states (ADOS) corresponding to the results presented in Figs.~\ref{fig:cfb}(a) and~\ref{fig:cfb}(b). 
Using the standard Green's function technique, ADOS can be defined as,
\begin{equation}
\rho(E) = -\dfrac{1}{N\pi}\textrm{Im} \left[ \textrm{Tr}\; {\bm G}(E) \right],
\label{eq:ADOS}
\end{equation} 
where ${\bm G}(E) = \left[ z^{+}{\bm I} - \bm{H} \right]^{-1}$ is the Green's function with 
$z^{+}=E+i\delta\ (\delta \rightarrow 0^{+})$, $N$ is the total number of sites in the system, 
and `$\textrm{Tr}$' denotes the trace of the Green's function ${\bm G}$. 
Using Eq.~\eqref{eq:ADOS}, we calculate the ADOS for our lattice model with a system size 
$L=50 \times 50$, $L$ being the number of unit cells. The results are shown in Figs.~\ref{fig:ADOS}(a) 
and~\ref{fig:ADOS}(b). Evidently, the presence of highly localized spiky states exactly at the FB energies 
confirms the appearance of the complete FB states in our lattice model.   
\begin{figure}[ht]
\includegraphics[clip,width=0.8\columnwidth]{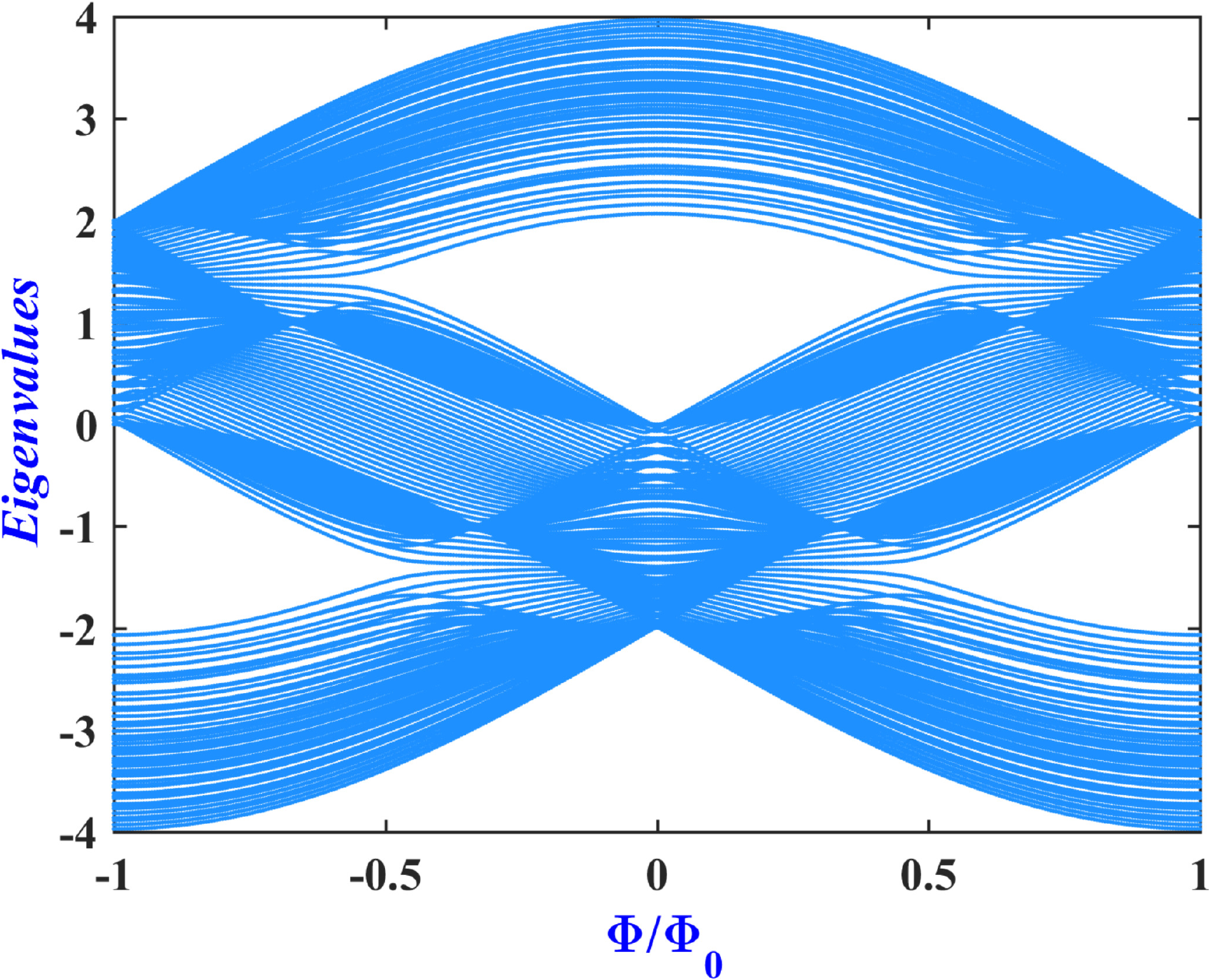}
\caption{Plot of the energy eigenvalue spectrum against the magnetic flux $\Phi$ 
(measured in units of the fundamental flux quantum $\Phi_{0}=hc/e$) for a finite 
system of size $L=10 \times 10$, $L$ being the number of unit cells.}
\label{fig:E-Phi}
\end{figure}

Before ending this section, we present the energy eigenvalue spectrum of the real-space Hamiltonian against 
the variation of the magnetic flux ($\Phi$) for a finite system with system size $L=10 \times 10$. 
This is exhibited in Fig.~\ref{fig:E-Phi}. This result gives us the flavor about the real-space 
energy spectrum of the system in presence of $\Phi$. From Fig.~\ref{fig:E-Phi}, we can clearly observe 
the formation of multiple bands and gaps in the spectrum as a function of $\Phi$. A variation in the value of 
the magnetic flux leads to band overlapping in spectrum, and the whole pattern is flux periodic. The spectrum 
will be more and more dense as we increase the system size, but the overall shape will remain the same.  
\section{Possible experimental realization of the model}
\label{expt}
In this section, we will discuss the possibility of an experimental realization of our lattice model using 
photonic waveguide structure. The femtosecond laser-writing technique along with the aberration-correction 
methods~\cite{mukherjee-prl2015} allow us for the precise fabrication of two-dimensional arrays of sufficiently 
deep single-mode waveguides. The advantage of such techniques over other photonic platforms is that the 
laser-writing parameters can be optimized to produce low propagation loss over a long distance implicating 
single-mode waveguides to operate at a particular wavelength. In addition to that, this method also gives us 
an efficient control over the inter-waveguide coupling strengths allowing us to explore different parameter regimes. 
Such techniques have been successfully implemented in recent times to accomplish experimental 
realization of flat bands in a Lieb photonic structure~\cite{vicencio-prl2015, mukherjee-prl2015}, and 
soon followed by other photonic lattice geometries~\cite{mukherjee-ol2015,longhi-ol2014,zong-oe2016, 
weimann-ol2016}.     
\begin{figure}[ht]
\includegraphics[clip,width=0.7\columnwidth]{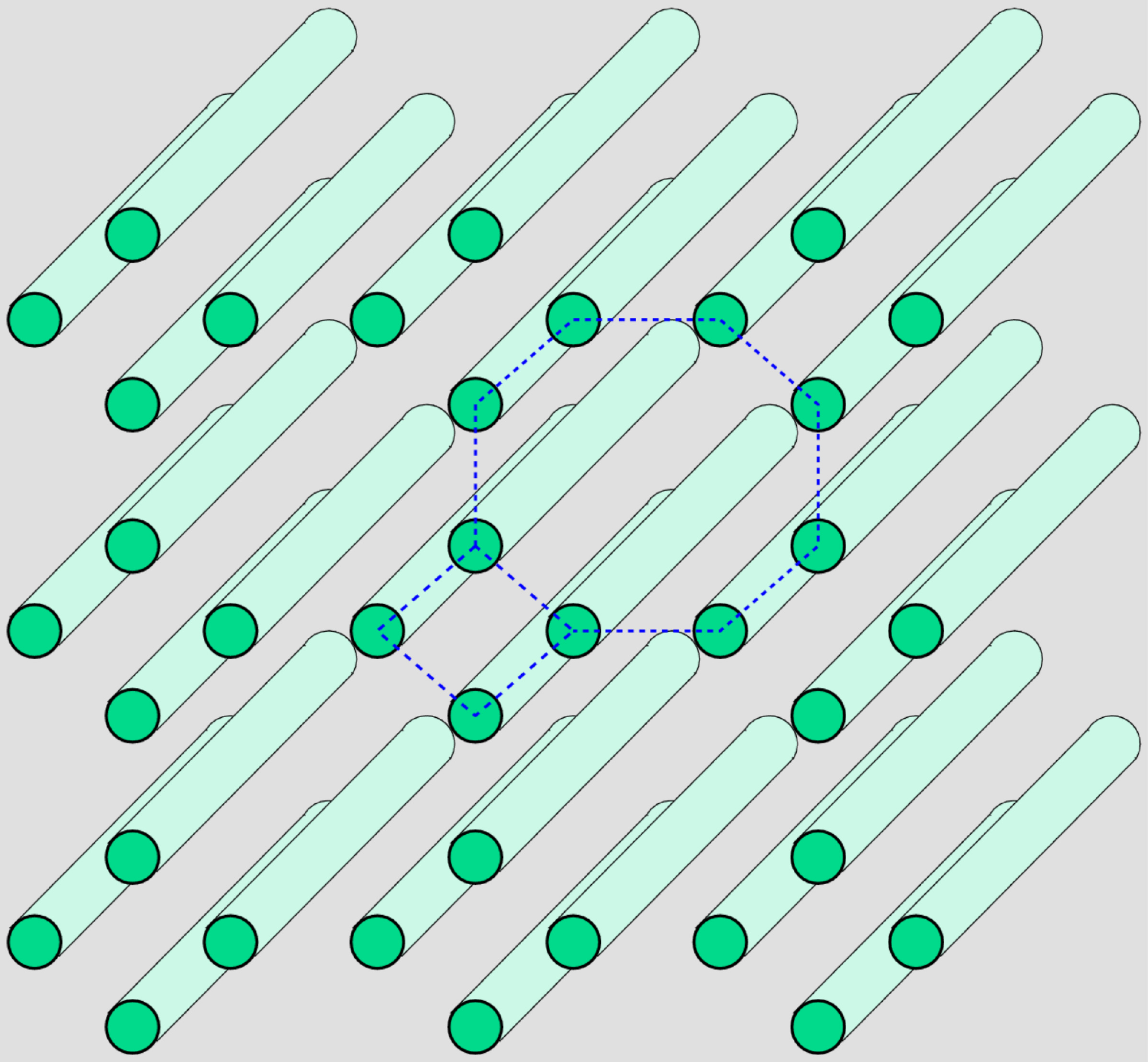}
\caption{Schematic representation of a possible proposed photonic waveguide 
network corresponding to the 2D lattice model depicted in Fig.~\ref{fig:lattice}. 
Each lattice site is substituted by single-mode waveguides to form the waveguide structure.}
\label{fig:waveguide}
\end{figure}

Considering the structural homology of our lattice model in comparison with other 2D lattice geometries 
such as Lieb or kagome structure, we strongly emphasize that our lattice structure can be fabricated 
expeditiously using photonic waveguides to study the formation of the flat bands and other related 
interesting properties. A schematic representation of such possible photonic waveguide structure 
corresponding to our lattice geometry has been displayed in Fig.~\ref{fig:waveguide}. The values of 
the system parameters for these photonic waveguide structures, such as lattice period, propagation distance, and 
operating wavelength are typically chosen in the range of 20-30 $\mu$m, 7-10 cm, and 500-800 nm, 
respectively~\cite{vicencio-prl2015, mukherjee-prl2015, mukherjee-ol2015, longhi-ol2014}. The 
exact value of these parameters may vary slightly depending on the experimental conditions for obtaining the flat bands. 
The effect of the external magnetic field can be simulated in a coupled waveguide network by incorporating 
a synthetic magnetic field through a proper longitudinal modulation of the propagation constants of the 
waveguides~\cite{longhi-ol2014}. The phenomenon of time-reversal symmetry breaking in circuit-QED based 
photon lattices has also been reported earlier~\cite{girvin-pra2010}. Such mechanisms could be helpful 
to accomplish our results in presence of an external magnetic field in an actual experimental setup using 
photonic waveguides. In addition to the advancement of the fundamental understanding of the physics of 
flat bands, the photonic flat band networks can also have technological importance in photonics, such 
as slow light propagation~\cite{baba-np2008}, where the suppression of the wave group velocity can 
provide enhancement of nonlinear effects, and promising solution for buffering and time-domain 
processing of optical signals. Our lattice model implemented using a photonic waveguide network can 
provide a potential platform to realize such useful devices in photonics. Apart from the photonic 
lattice structure, there has been a remarkable technological advancement in developing artificial 
lattice structures using ultracold atomic condensates in optical lattices~\cite{bloch-rmp2008}. 
This can also be utilized to engineer our lattice model in experiments, and study its interesting 
novel properties in a very controllable and clean environment without the presence of the impurities 
appearing in a typical solid state system.     
\section{Summary and future outlook}
\label{summary}
In this paper, we have investigated the energy spectrum of a tight-binding diamond-octagon lattice model 
containing flat band states. We have perceived that for a suitable combination of the hopping parameters 
and an external magnetic flux, it is possible to realize nearly flat band states with nonzero Chern numbers 
for this model. The presence of such bands in the energy spectrum may lead to a very interesting scenario 
for incarnating strongly correlated electronic states with nontrivial topological properties. For a fractional 
filling in the ground state of our system, one can envision the fractional quantum Hall physics in a lattice 
model. In addition to that, we have also revealed the existence of perfectly flat band states in our lattice model, 
forming compact localized states. The calculation of the density of states and the wavefunction amplitude 
distribution on lattice sites corroborate the formation of the compact localized states corresponding to the 
flat band states in our model. Our work has put forward a simple example of a 2D tight-binding lattice model 
to understand certain important flat band physics in a lattice system. We believe that the experimental 
realization of our model using untracold atoms in optical lattices is definitely on the card, and may unfold 
interesting topological phases of matter. One can also fabricate a photonic diamond-octagon lattice using 
single-mode photonic waveguides controlled by femtosecond laser pulses~\cite{mukherjee-prl2015, 
mukherjee-ol2015} to study the photonic flat bands in such a lattice model. The investigation of the 
robustness of these flat band states encountered in our model under different perturbing effects such 
as spin-orbit interaction, disorder etc. could be an open direction for further exploration.       
\begin{acknowledgments}
The author gratefully acknowledges a postdoctoral scholarship and the 
computational facility provided by MPIPKS. The author also thanks 
K. Saha and B. Roy for helpful discussions, and N. S. Srivatsa for 
a critical reading of the manuscript. 
\end{acknowledgments}

\end{document}